\documentclass[12pt]{article}
\usepackage{amsthm}
\usepackage{amsmath}
\usepackage{natbib}
\usepackage[colorlinks,citecolor=blue,urlcolor=blue,filecolor=blue,backref=page]{hyperref}
\usepackage{graphicx}
\usepackage{amsmath}
\usepackage{graphicx,psfrag,epsf}
\usepackage{enumerate}
\usepackage{natbib}
\usepackage{graphicx}			
\usepackage{amssymb,amsmath,amsthm,bm} 
\usepackage{mathtools}
\usepackage{commath}
\usepackage[english]{babel}
\usepackage{hyperref}
\usepackage{threeparttable}
\usepackage{bbm}
\usepackage{enumerate}
\usepackage{float}
\usepackage{bbm}
\usepackage{graphicx} 
\usepackage{booktabs} 
\usepackage{algorithm,algcompatible,amsmath}
\usepackage{stackrel}

\usepackage{rotating}
\usepackage{booktabs}  
\usepackage{siunitx}  
\usepackage{caption}
\usepackage{bm}
\setcitestyle{authoryear,open={(},close={)}}

\usepackage{tikz}
\usetikzlibrary{fit,positioning}
\usepackage{tikz}
\usetikzlibrary{bayesnet}
\usetikzlibrary{arrows}

\newcommand{\blind}{0}
\renewcommand{\vec}[1]{\mathbf{#1}}

\DeclareMathOperator*{\argmax}{arg\,max}

\begin{document}

\def\spacingset#1{\renewcommand{\baselinestretch}%
{#1}\small\normalsize} \spacingset{1}


\if0\blind
{
  \title{\bf Finite Mixtures of ERGMs for Modeling Ensembles of Networks \thanks{This work was supported in part by NSF award DMS-1361425; corresponding author Carter T. Butts (buttsc@uci.edu)}}
  \author{Fan Yin\hspace{.2cm}\\
    Department of Statistics, University of California, Irvine\\
    Weining Shen \\
    Department of Statistics, University of California, Irvine \\
    Carter T. Butts \\
    Department of Statistics, Sociology, Computer Science, \\
    and EECS, and Institute for Mathematical Behavioral Sciences, \\
    University of California, Irvine}
  \maketitle
} \fi
\if1\blind
{
  \bigskip
  \bigskip
  \bigskip
  \begin{center}
    {\LARGE\bf Title}
\end{center}
  \medskip
} \fi

\bigskip
\begin{abstract}
Ensembles of networks arise in many scientific fields, but there are few statistical tools for inferring their generative processes, particularly in the presence of both dyadic dependence and cross-graph heterogeneity. To fill in this gap, we propose characterizing network ensembles via finite mixtures of exponential family random graph models, a framework for parametric statistical modeling of graphs that has been successful in explicitly modeling the complex stochastic processes that govern the structure of edges in a network. Our proposed modeling framework can also be used for applications such as model-based clustering of ensembles of networks and density estimation for complex graph distributions. We develop a Metropolis-within-Gibbs algorithm to conduct fully Bayesian inference and adapt a version of deviance information criterion for missing data models to choose the number of latent heterogeneous generative mechanisms. Simulation studies show that the proposed procedure can recover the true number of latent heterogeneous generative processes and corresponding parameters. We demonstrate the utility of the proposed approach using an ensemble of political co-voting networks among U.S. Senators.  
\end{abstract}

\noindent%

\spacingset{1.45}
\section{Introduction}
\label{sec:intro}

Data involving ensembles of networks - that is, multiple independent networks - arise in various scientific fields, including sociology \citep{slaughter2016multilevel, stewart2019multilevel}, neuroscience \citep{simpson2011exponential, obando2017statistical}, molecular biology \citep{unhelkar2017structure,grazioli2019comparative}, and political science \citep{moody2013portrait} among others. Typically, ensembles of networks represent the action of multiple generative processes, with different processes being prominent in different settings. A reasonable starting point for analysis of such data is to posit that this variation can be represented in terms of discrete set of subpopulations, such that the networks drawn from any given subpopulation tend to be produced by similar generative processes. Given a set of potential generative models, one would then like to identify the subsets of networks drawn from a particular subpopulation, or a probabilistic mixture of multiple subpopulations. It is natural to view this as a hierarchical finite mixture problem, with the base distributions being parametric distributions on graphs. 
As a plausible approximation to the underlying data generating process, the hierarchical finite mixture framework also provides a flexible approach for predictive modeling of ensemble of networks. If one seeks to predict graph structures drawn from a heterogeneous (super)population learned from observed data, one needs to average over the possible generative processes that might end up producing the observation that one wants to predict. Such a view is similar in spirit to model averaging techniques \citep{hoeting1999bayesian,hjort2003frequentist}, especially if interpreted in terms of a hierarchical problem in which we seek to predict an outcome of interest (e.g., co-voting prevalence among U.S. senators) by first predicting network structure and then predicting the behavior of a process on that network. In that setting, if it turned out that there were $k$ types of possible network formation processes and we did not know which one ours happened to be, we would certainly want to average across the types. 

There is a growing body of literature on the analysis of ensembles of networks. This includes work on discriminative analysis of networks via distance or similarity measures \citep[e.g.][]{banks.carley:joc:1994,butts.carley:cmot:2005,fitzhugh.et.al:alcr:2015}, which can be broadly viewed as mapping the ensemble of interest into some high-dimensional space (e.g., the Hamming space of graphs), and then employing standard multivariate analysis techniques (e.g., hierarchical clustering, multidimensional scaling) to seek an informative low-dimensional approximation.  Other approaches work with user-selected graph statistics, either directly \citep[e.g.][]{przulj:b:2007,sweet2019clustering} or by e.g., modeling quantiles of the observed statistics relative to a reference distribution to control for size and density effects \citep{butts2011bayesian}. As such, these approaches do not attempt to provide generative models for the networks within the ensemble, though they may in some cases provide generative models for summary statistics (e.g., predicting the conditional uniform graph quantile for the transitivity of a new graph drawn from the same ensemble).

In the category of generative models for complex networks, a common approach is to employ multilevel models with exponential random graph models (ERGMs, a general family of parametric models for networks \citep[see, e.g.][for a review]{robins2007recent}), as base distributions.  \citet{faust2002comparing} introduced both multivariate meta-analysis of ERGM parameters from a common model family (fit to an ensemble of graphs) and predicted conditional edge probabilities from the generative base models as tools for leveraging ERGMs to compare networks. More elaborate meta-analytic procedures and hierarchical models for single populations of networks were subsequently developed by, among others, \citet{zijlstra2006multilevel, slaughter2016multilevel, mcfarland.et.al:asr:2014, butts2017baseline}, and \citet{stewart2019multilevel}. Nonparametric models (e.g., latent space or block models) have also been employed for studying sets of networks, e.g. hierarchical mixed membership stochastic blockmodels for multiple networks \citep{sweet2014hierarchical}. In general, those methods have either not posited a generative model for the parameters of the base distribution (as in descriptive meta-analytic approaches), have not attempted to jointly estimate population-level and network-level parameters (as in conventional meta-analysis), or have assumed a simple hierarchical form in which coefficients are taken to be drawn from a simple population distribution (often Gaussian) with common mean and variance.  The latter work well for homogeneous (super)populations; but when the network ensemble reflects higher levels of heterogeneity, more structure is required. In contrast, work such as that of \citet{durante2018bayesian,lehmann2019inferring} explicitly considers heterogeneity within graph subpopulations, but assumes that the subpopulation labels are observed. Joint modeling of population-level and network-level parameters where subpopulation memberships are unknown, or where the true generative process otherwise involves a mixture of graph distributions, has remained an open problem to date in the ERGM context.

In this paper, we propose using a mixture of ERGMs to model the generative process of ensembles of networks in which the group labels are not available, under the general framework of finite mixture models \citep{mclachlan1988mixture, fraley2002model, bouveyron2019model}. Such a formulation provides a useful probabilistic interpretation of the results and allows for convenient statistical inference; we note that related approaches have proven to be efficacious for modeling structure \emph{within} networks \citep[e.g.][]{salter2015role,schweinberger2015local,snijders1997estimation}. Recent work on using mixtures of network models with dyadic independence property (e.g., a priori stochastic blockmodel, $p_1$ model) for modeling multiple network observations \citep{signorelli2019model} can encounter difficulties when the observed networks exhibit strong dyadic dependence, which is often the case for real-world networks. We develop a Metropolis-within-Gibbs algorithm to perform Bayesian inference for the proposed model, with both the subpopulation  assignments and the ERGM parameters in the subpopulations being estimated simultaneously. Given that our primary focus is to develop a practical procedure that can obtain meaningful subpopulations, we employ a pseudo-likelihood approximation to the ERGM likelihood for efficient computation; while we show here that this approach can work well, more advanced MCMC techniques can also be deployed to obtain more accurate estimates when the interest lies mainly in the inference of subpopulations-specific parameters.  (It is also possible to use the pseudo-likelihood when updating subpopulation assignment parameters and then use high-accuracy MCMC-based likelihood calculations to update subpopulation-specific parameters, offering additional options for speed/accuracy tradeoffs.) We approach the problem of choosing number of subpopulations from a model selection perspective, using a version of deviance information criterion.

The remainder of this paper is structured as follows. In section \ref{sec:ERGMs} we briefly introduce the exponential-family random graph models (ERGMs) and common estimation techniques. Section \ref{sec:Mixture_of_ERGMs} describes the idea of mixtures of ERGMs, along with our estimation algorithms and our proposed method for selecting the number of subpopulations. Section \ref{sec:Simulation} presents simulation studies showing that the proposed method can accurately recover the true subpopulation assignment and model parameters. Section \ref{sec:Case_study} shows the results of our method applied to a political co-voting data analysis. Section \ref{sec:Conclusion} concludes with a discussion. 

\section{Exponential-family Random Graph Models (ERGMs)}
\label{sec:ERGMs}
In recent years, ERGMs have found applications in empirical research in a wide range of scientific fields. Recent examples include the study of large friendship networks \citep{goodreau2007advances}, genetic and metabolic networks \citep{saul2007exploring}, disease transmission networks \citep{groendyke2012network}, conflict
networks in the international system \citep{cranmer2011inferential}, the
structure of ancient networks in various of archaeological settings \citep{amati2019framework}, the structural comparison of protein structure networks \citep{grazioli2019comparative}, the effects of functional integration and functional segregation in brain functional connectivity networks \citep{simpson2011exponential,sinke2016bayesian,obando2017statistical}, and the impact of endogenous network effects on the formation of interhospital patient referral networks \citep{caimo2017bayesian}. While addressing very
different problems in different empirical settings, what these studies have in common is a clear
methodological commitment to modeling network mechanisms directly via parametric effects, rather than just attempting to ``control for'' unspecified dependence among the observations (e.g., via latent structure).  The ability to provide generative and interpretable models of complex network structure is an important asset of this approach, which we leverage here in the context of graph ensembles.

\subsection{Definition and Estimation}
\label{subsec:ERGM_def}
Exponential-family random graph models (ERGMs) \citep{holland1981exponential, frank1986markov, snijders2006new, hunter2006inference}, also known as $p$-star models \citep{wasserman1996logit}, are a family of parametric statistical models developed for explicitly modeling the complex stochastic processes that govern the formation of edges among pairs of nodes in a network. We introduce them first in the single-network case. Consider the set of nodes in the network of interest, $\vec{V}$, and let $|\vec{V}| = n$ be its cardinality, i.e. the number of nodes in the network.  We represent the network's structure via an order-$n$ random adjacency matrix $\vec{Y}$, in which each element takes $1$ or $0$ representing the presence or absence of a tie between incident nodes. Letting $\mathcal{Y}_{n}$ be the set of all possible network configurations on $n$ nodes, we write the probability mass function (pmf) of $\vec{Y}$ taking a particular configuration $\vec{y}$ in the form of a discrete exponential family as
\begin{equation} 
\label{eq:ERGM}
\mathbb{P}_{\bm{\eta}}( \vec{Y} = \vec{y}|\vec{X}; {\bm{\theta}}) = \exp \bigg( \bm{\eta}( \bm{\theta})^{\intercal} \vec{g(y;\vec{X})} - \psi_{\vec{g}, \bm{\eta} ,\vec{X}, \mathcal{Y}_{n}}( \bm{\theta} ) \bigg) h(\vec{y}),  \quad \vec{y} \in \mathcal{Y}_{n},
\end{equation}

where $\bm{\theta} = (\theta_{1}, \cdots, \theta_{q}) \in \mathbb{R}^{q}$ is a vector of (curved) model parameters, mapped to the natural parameters by $ \bm{\eta}(\bm{\theta}) = ( \eta_{1}(\bm{\theta}), \cdots,  \eta_{p}(\bm{\theta})  ) \in\mathbb{R}^{p}$. The natural parameters $\bm{\eta}$ may depend on the sizes of the networks and may be non-linear functions of a parameter vector $\bm{\theta}$. The user-defined sufficient statistics $\vec{g} : \mathcal{Y}_{n} \rightarrow \mathbb{R}^{p}$ may incorporate fixed and known covariates $\vec{X}$ that are measured on the nodes or dyads. The sufficient statistics incorporate network features of interests that are believed to be crucial to the social process which had given rise to it \citep[see, e.g.,][]{morris2008specification}. Here $h$ defines the reference measure for the model family; often chosen to be the counting measure on $\mathcal{Y}_n$ for unvalued graphs with fixed $n$, other reference measures can make more sense in different settings.  As discussed below, we employ a sparse graph reference that leads to a mean degree that is asymptotically constant in $n$.  Finally, the normalizing factor $\psi_{\vec{g}, \bm{\eta}, \vec{X}, \mathcal{Y}_{n}}( \bm{\theta} ) = \log \sum_{\vec{y'} \in \mathcal{Y}_{n} } \exp\left\{ \bm{\eta}( \bm{\theta})^\intercal \vec{g(\vec{y'};\vec{X})} \right\} h(\vec{y'})$ ensures that \eqref{eq:ERGM} sums up to 1 over the support $\mathcal{Y}_{n}$. To make notations simpler, we also assume that $\vec{V}$ is implicitly absorbed into $\vec{X}$.

Exact evaluation of the normalizing factor involves integrating an extremely rough function over all possible network configurations ($2^{n \choose 2}$ non-negative terms for an undirected network of size $n$). This cannot be done by brute force except for trivially small graphs, and the roughness of the underlying function precludes simple Monte Carlo strategies; thus, alternative approaches that approximate or avoid this calculation are of substantial interest \citep[see][for a review]{hunter2012computational}. To date, the most frequently used approaches include: maximum pseudo-likelihood estimation (MPLE; \citet{besag1974spatial}) adapted by \citet{strauss1990pseudolikelihood}; Markov Chain Monte Carlo MLE (MCMC MLE; \citet{geyer1992constrained}) by \citet{handcock2003assessing, hunter2006inference}; Stochastic approximation (SA; \citet{robbins1951stochastic,pflug1996optimization}) by \citet{snijders2002markov}; and fully Bayesian inference based on approximate exchange algorithm \citep{caimo2011bayesian}. Recent developments on ERGM estimation have concentrated on: (1) finding better initial values for simulation-based MLE, including the \emph{partial stepping} technique \citep{hummel2012improving} and \emph{contrastive divergence} (CD,\citet{hinton2002training})-based techniques adapted to ERGMs by \citet{krivitsky2017using}; and (2) more accurate tractable approximations to ERGM likelihood than pseudo-likelihood, such as the adjusted pseudo-likelihood \citep{bouranis2017efficient,bouranis2018bayesian} for fast Bayesian inference. Despite the computational challenges, these and related strategies have made ERGM inference practical for well-posed model families (e.g., see \citep{schweinberger2019exponential} for a recent review). 

\subsection{Size-adjusted parameterizations}
It is worth noting that the behavior of Eq.~\eqref{eq:ERGM} across $n$ is highly dependent on the choice of reference measure, $h$.  In particular, the counting measure - while a mathematically convenient choice - implicitly sets the base distribution of the network to be the uniform distribution on $\mathcal{Y}_n$, and has the side effect of generating graphs whose densities are \emph{ceteris paribus} constant in $n$.  When network size varies, this is not always realistic: in many networks, mean degree is approximately constant in $n$, implying that density must scale as $n^{-1}$.  To correct for this, \citet{krivitsky2011adjusting} propose the reference measure $h(\vec{y})=n^{-M(\vec{y})}$, where $M$ is the edge count.  This is equivalent to adding a size-dependent offset of $-\log n$ to the natural parameter associated with the edge count, i.e.
\begin{equation}
\label{eq:krivitsky_offset}
\eta_{1}(\bm{\theta}) = \theta_{1} - \log n,
\end{equation}
\noindent where $\theta_{1} \in \mathbb{R}$ is a parameter that does not depend on the network size.  In the present work, we employ the \emph{Krivitsky reference measure} as above, although other size-adjusted parameterizations are also possible \citep[e.g., ][]{butts2015flexible, kolaczyk2015question}.

\section{Finite mixtures of ERGMs}
\label{sec:Mixture_of_ERGMs}
We assume a population of networks $(\vec{Y}^{(1)},\vec{V}^{(1)},\vec{X}^{(1)}),\ldots,(\vec{Y}^{(m)},\vec{V}^{(m)},\vec{X}^{(m)})$, where $\vec{Y}^{(i)}$ is a graph structure on vertex set $\vec{V}^{(i)}$ with covariate set $\vec{X}^{(i)}$.  Our interest is in modeling $\vec{Y}^{(1)},\cdots, \vec{Y}^{(m)}$ given $(\vec{V}^{(1)},\vec{X}^{(1)}),\cdots,(\vec{V}^{(m)},\vec{X}^{(m)})$, where it will be assumed that the respective graph structures are conditionally independent given the generative process, vertex sets, and covariates.

\subsection{Model}
\begin{figure}
\begin{center}
\includegraphics[width=0.5\textwidth]{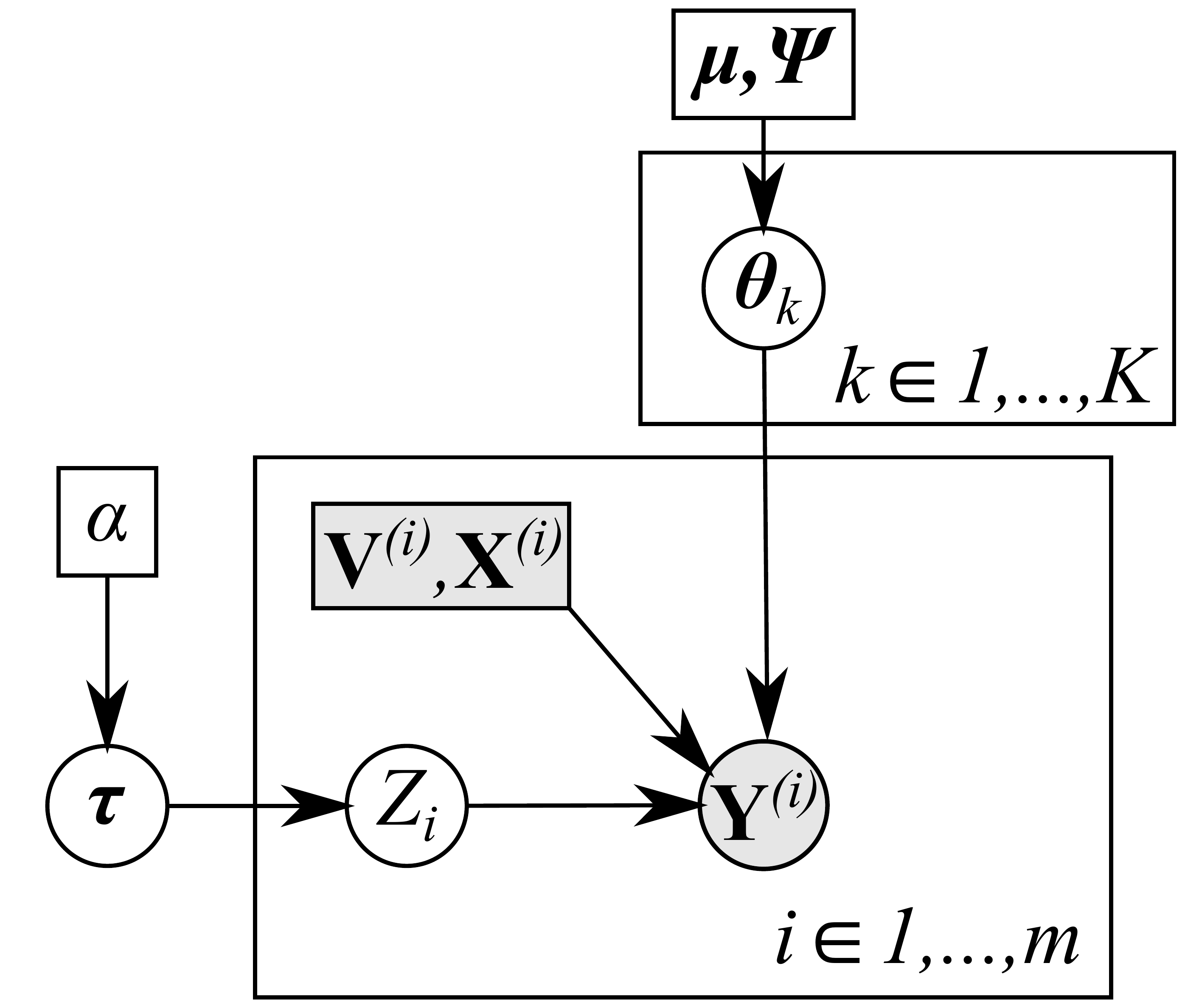}
\caption{Structure of the graph mixture model.  Random quantities are depicted within circles, fixed quantities within rectangles; observables are shaded.\label{fig:mixture_model_graph}}
\end{center}
\end{figure}

We model the generative process for the network ensemble as a finite mixture, with each mixture component (equivalently, subpopulation, or ``cluster'') being an ERGM distribution with cluster-specific parameters. (See Figure~\ref{fig:mixture_model_graph}.)  Given $K$ clusters, the {\it a priori} probability for a network to belong to cluster $k$ is $\tau_{k}$ for $k=1,2,\cdots,K$, and the probability law governing the formation of the network in group $k$ is parameterized by Eq.~\eqref{eq:ERGM} with cluster-specific parameter vector $\bm{\theta}_{k} \in \mathbb{R}^{q_{k}}$ and cluster-specific mapping to the natural parameters $\bm{\eta}_{k}(\bm{\theta}_{k}) = ( \eta_{k,1}(\bm{\theta}_{k}), \cdots,  \eta_{k,p_{k}}(\bm{\theta}_{k})) \in \mathbb{R}^{p_{k}}$. For notational simplicity, we omit the subscripts $\bm{\eta}_{k}$'s for the remainder of the paper. 

More specifically, the marginal likelihood for network $\vec{Y}^{(i)}$, with $|\vec{V}^{(i)}| \equiv n_{i}$, takes the following form
\begin{equation} 
\label{eq:mixture_ERGM}
\mathbb{P}(\vec{Y}^{(i)} = \vec{y}^{(i)} | \vec{X}^{(i)}; \bm{\tau}, \underline{\bm{\theta}}) = \sum_{k=1}^{K} \tau_{k} \exp \bigg( \bm{\eta}_{k}(\bm{\theta}_{k})^{\intercal} \vec{g}_{k}(\vec{y}^{(i)};\vec{X}^{(i)}) - \psi_{\vec{g_{k}},\bm{\eta}_{k}, \vec{X}^{(i)},\mathcal{Y}_{n_{i}}}(\bm{\theta}_{k})  \bigg) h_i(\vec{y}^{(i)}),  \vec{y}^{(i)} \in \mathcal{Y}_{n_{i}}
\end{equation}

where $\bm{\tau} = (\tau_{1}, \cdots, \tau_{K})$ and $\underline{\bm{\theta}} = (\bm{\theta}_{1},\cdots,\bm{\theta}_{K})$ are the model parameters, and the former satisfies the constraint $\sum_{k=1}^{K} \tau_{k} = 1, \tau_k \geq 0$ for $k=1,\ldots,K$. 

The ensemble of networks consists of $m$ independent observations $ \underline{\vec{y}} = (\vec{y}^{(1)},\cdots,\vec{y}^{(m)})$ with fixed covariate set $\underline{\vec{X}} = (\vec{X}^{(1)},\cdots,\vec{X}^{(m)})$ and fixed vertex set $\underline{\vec{V}} = (\vec{V}^{(1)},\cdots,\vec{V}^{(m)})$, and hence the joint likelihood is
\begin{equation} 
\label{eq:mixture_ERGM_joint}
\mathbb{P}(\underline{\vec{Y}} = \underline{\vec{y}}  | \underline{\vec{X}}; \bm{\tau}, \underline{\bm{\theta}}) = \prod_{i=1}^{m} \bigg[ \sum_{k=1}^{K} \tau_{k} \exp \bigg( \bm{\eta}_{k}(\bm{\theta}_{k})^{\intercal} \vec{g}_{k}(\vec{y}^{(i)};\vec{X}^{(i)}) - \psi_{\vec{g_{k}},\bm{\eta}_{k}, \vec{X}^{(i)}, \mathcal{Y}_{n_{i}}}(\bm{\theta}_{k})  \bigg) h_i(\vec{y}^{(i)}) \bigg],
\end{equation}

where we have absorbed the support constraint into the reference measure.

To facilitate statistical inference, we consider the representation of \eqref{eq:mixture_ERGM_joint} from a latent variable perspective. Let $Z_{i}, i=1,\cdots,m$ be latent variables following a categorical distribution with $K$ values and probability parameter $\bm{\tau}$, such that $Z_{i} = k$ if $\vec{Y}^{(i)}$ belongs to cluster $k$.  We may then treat $\vec{Y}^{(i)}$ as arising from a process in which $Z_i$ is first drawn from $\mathrm{Categorial}(\bm{\tau})$, and $\vec{Y}^{(i)}$ is then drawn from the ERGM distribution corresponding to cluster $Z_i$.  While one could allow the reference measure to also vary by cluster, we focus on the case of ERGMs specified relative to the Krivitsky reference measure if the sizes of the networks vary. 

\subsection{Bayesian estimation}
Bayesian estimation is a natural choice for parameter inference here, since (1) it is more robust to initialization and less prone to converge to local minima than maximum likelihood; (2) interval estimation is straightforward and does not rely on the assumption of approximate normality; and (3) it provides principled answers in fixed-$n,m$ settings. Our strategy is to employ Metropolis-within-Gibbs sampling to obtain MCMC samples from the joint posterior distribution of $\underline{\bm{\theta}}$ and $\bm{\tau}$.  

We specify prior distributions for the parameters as follows, 

$$ \bm{\tau} \sim \text{Dirichlet}(\bm{\alpha}), $$

$$ \bm{\theta}_{k}  \overset{i.i.d.}{\sim} \text{MVN}_{p}(\bm{\mu}, \Psi  ), \ \ \ k=1,\cdots,K, $$

where $ \bm{\alpha} = (\alpha_{1}, \cdots, \alpha_{K})$, $\bm{\mu}$ and $\Psi$ are hyper-parameters to be specified by the user. For typical use cases, a reasonable choice of hyperparameters are  $\alpha_{1} = \ldots = \alpha_K = 3$, which puts low probability on any group being extremely small, and $\Psi = 25 I_{p}$, which is fairly flat over the typical range of variation for common parameterizations. A convenient choice of $\bm{\mu}$ is $\vec{0}$, but this can be problematic because it will rarely be true that we want to shrink the edge parameter (which governs density) towards 0. It can hence be important to incorporate empirical knowledge into the specification of $\bm{\mu}$; in particular, set the hyperparameter associated with edge term to be negative (e.g., $-\log n$) when modeling social networks under the counting measure, as most social networks are sparse.  Under the Krivitsky reference measure, using the log of the \emph{a priori} expected degree (based either on theory or analysis of similar data sets) is an appropriate choice.

As noted, we perform posterior inference via MCMC. Our algorithm iterates over the model parameters $(\underline{\bm{\theta}}, \bm{\tau})$ with the priors given above, and the latent variables $\vec{Z} = (Z_{1}, \cdots, Z_{m})$.  Where possible we sample from the full conditional posterior distributions; otherwise we use Metropolis-Hastings steps. 

\begin{algorithm}
    \caption{Metropolis-within-Gibbs sampler for the ERGM mixture model \label{alg:alg1}}
  \begin{algorithmic}[1]
     \STATE \textbf{Initialization}: Set $\bm{\tau}^{0}$, $\underline{\bm{\theta}}^{0}$ and $\bm{Z}^{0}$ to initial values (e.g., prior means). 
    
    \FOR{$t = 1,2,\cdots,T $}
    \STATE Generate $Z_{i}^{t}$ ($i=1,\cdots,m$, $k=1,\cdots,K$) from \newline
            \indent $\mathbb{P}( Z_{i}^{t} = k | \eta_{k}^{t-1}, \bm{\theta}_{k}^{t-1}, \vec{y}^{(i)}   ) \propto \eta_{k}^{t-1} \mathbb{P}(\vec{y}^{(i)} | \vec{X}^{(i)}; \bm{\theta}_{k}^{t-1} )$
    \STATE Compute $\nu_{k}^{t} = \sum_{i=1}^{m} \mathbbm{1}_{Z_{i}^{t} = k}$; $k=1,\cdots,K$
    \STATE Generate $\bm{\tau}^{t}$ from $\text{Dirichlet}(\alpha_{1} + \nu_{1}^{t},\cdots, \alpha_{K} + \nu_{K}^{t} )$ 
      \FOR{$k=1,\cdots,K$}
      \STATE Propose $\bm{\theta}_{k}^{'} \sim q(\cdot | \bm{\theta}_{k}^{t-1})$
      \STATE Accept $\bm{\theta}_{k}^{'}$ with probability equal to \newline
      \indent $\frac{ \pi( \bm{\theta}_{k}^{'} ) \prod_{ Z_{i}^{t}=k } \mathbb{P}(\vec{y}^{(i)} | \vec{X}^{(i)}; \bm{\theta}_{k}^{'}) q( \bm{\theta}_{k}^{t-1} | \bm{\theta}_{k}^{'})   }{   \pi( \bm{\theta}_{k}^{t-1} ) \prod_{ Z_{i}^{t}=k } \mathbb{P}(\vec{y}^{(i)} |\vec{X}^{(i)}; \bm{\theta}_{k}^{t-1}) q(  \bm{\theta}_{k}^{'}|\bm{\theta}_{k}^{t-1}) } $ \label{eq:MH_ratio}
      \ENDFOR
     \ENDFOR
  \end{algorithmic}
\end{algorithm}

The proposal distribution $q(\cdot | \bm{\theta})$ in the Metropolis step is set by the user to achieve good performance of the algorithm. On the basis of some experimentation, we use the symmetric proposal $\mathcal{N}(\bm{\theta}, \sigma^{2} I_{q})$, where $\sigma = 0.05$. At each MCMC iteration, we permute the labels to impose ordering constraints on the first common element of the parameter vectors (e.g., total number of edges), $\theta_{11} < \theta_{21} < \cdots < \theta_{K1} $ for model identifiability purposes. Simulation studies and case studies show that the ordering constraints can work well, though other post-processing techniques (e.g., Kullback-Leibler relabeling algorithm \citep{stephens2000dealing} and Pivotal Reordering algorithm \citep{marin2005bayesian}, etc.), can be used depending on practitioners' preference. 

To deal with the intractability of $\mathbb{P}( \vec{y}^{(i)} | \vec{X}^{(i)}; \bm{\theta})$, there are at least three possible solutions in ERGM literature:

\begin{itemize}
    \item Work with a tractable approximation in the place of ERGM likelihood, e.g. pseudo-likelihood \citep{strauss1990pseudolikelihood}, fully adjusted pseudo-likelihood \citep{bouranis2018bayesian}, or other composite likelihoods \citep{austad2010deterministic,asuncion2010learning}.
      \item Use importance sampling to approximate the ERGM likelihood \citep{koskinen2004bayesian, koskinen2008linked}.
      \item Use auxiliary-variable based MCMC algorithms to eliminate the intractable normalizing factor in ERGM likelihood \citep{caimo2011bayesian}. 
\end{itemize}

In fact, updating $\bm{\theta}_{k}$'s using the Metropolis-Hastings ratio in \eqref{eq:MH_ratio} is a \emph{doubly-intractable} problem, which can be approached using various advanced MCMC techniques \citep[see][for a review]{park2018bayesian}. However, these advanced techniques all require simulating networks from ERGMs at each MCMC iteration to approximate the true likelihood Eq.~\eqref{eq:ERGM}, which can be expensive for large networks. When the major goal is clustering instead of estimation on cluster-specific parameters, we propose to work with the most common form of tractable approximation, the pseudo-likelihood, in which the full likelihood of each network is approximated by a product of full conditional distributions of edge variables $y_{ij}$ in $\vec{y}$, 

\begin{equation}
\label{eq:ERGM_PL}
f_{PL}(\vec{y} | \vec{X}; \bm{\theta}) = \prod_{(i,j) \in \mathcal{D}} \mathbb{P}(y_{ij} | y_{-ij}; \vec{X}; \bm{\theta}) = \prod_{(i,j) \in \mathcal{D}} \frac{1}{1 + \exp\left\{-\bm{\eta}(\bm{\theta})^{\intercal} \Delta_{i,j} \vec{g}(\vec{y};\vec{X})  \right\}},
\end{equation}

where $\Delta_{i,j} \vec{g}(\vec{y}; \vec{X}) = \vec{g}(y_{ij}^{+};\vec{X}) - \vec{g}(y_{ij}^{-};\vec{X})$ are the so-called \emph{change statistics} associated with the dyad $(i,j)$, representing the change in sufficient statistics when $y_{ij}$ is toggled from 0 ($y_{ij}^{-}$) to 1 ($y_{ij}^{+}$) with the rest of the network remaining unchanged; $\mathcal{D}$ denotes the set of all pairs of dyads. For directed networks, $\mathcal{D} = \{ (i,j) | i,j \in \mathcal{N}, i \neq j  \}$, while for undirected networks, $\mathcal{D} = \{ (i,j) | i,j \in \mathcal{N}, i < j  \}$. In the frequentist paradigm, maximizing \eqref{eq:ERGM_PL} gives the so-called MPLE, which is relatively fast, algorithmically convenient, and able to provide approximate parameter estimates for even badly-specified models. While empirical observations show that MPLE can cause bias and underestimate standard errors \citep{van2009framework} (especially for models with strong dyadic dependence), it has been the default choice for initialization of MCMC-MLE algorithms. There is also promising work on using bootstrapped MPLE to construct confidence intervals \citep{schmid2017exponential} for large and sparse networks, as the MPLE is usually close to MLE in such cases \citep{desmarais2010consistent}.  Similar logic has motivated the use of Bayesian bootstrap estimation based on ``pseudo-MAP'' estimates using the PL approximation to the likelihood \citep{grazioli2019comparative}.

\subsection{Choosing the number of clusters}
\label{subsec:Clusters}
We recast the problem of choosing the number of clusters as a model selection problem, as different numbers of clusters result in distinct statistical models. Therefore, we use a version of the observed \textit{deviance information criteria} (DIC) introduced by \citet{celeux2006deviance}, which is an extension of the original DIC \citep{spiegelhalter2002bayesian} to models with latent variables. Given posterior draws $\bm{\tau}^{l},  \underline{\bm{\theta}}^{l} = (  \bm{\theta}_{1}^{l}, \cdots,  \bm{\theta}_{K}^{l})$ and observed ensemble of networks $\underline{\vec{y}} = (\vec{y}^{(1)}, \cdots, \vec{y}^{(m)})$, the observed DIC is defined by

\begin{equation}
\label{eq:DIC3}
DIC_{K} = -4 \mathbb{E}_{\underline{\bm{\theta}}}[ \log \mathbb{P}( \underline{\vec{y}} |  \underline{\vec{X}}; \underline{\bm{\theta}}) | \underline{\vec{y}}  ] + 2\log \hat{\mathbb{P}}(  \underline{\vec{y}}  | \underline{\vec{X}};  \underline{\bm{\theta}}),  
\end{equation}

where 

$$ \hat{\mathbb{P}}(  \underline{\vec{y}}  | \underline{\vec{X}}; \underline{\bm{\theta}}) = \prod_{i=1}^{m} \hat{\mathbb{P}}(\vec{y}^{(i)} | \vec{X}^{(i)}; \underline{\bm{\theta}}) = \prod_{i=1}^{m} \bigg( \frac{1}{m} \sum_{l=1}^{L} \sum_{k=1}^{K} \tau_{k}^{l} \mathbb{P}(\vec{y}^{(i)} | \vec{X}^{(i)}; \bm{\theta}_{k}^{(l)}  )  \bigg),  $$ 

and

$$ \mathbb{E}_{\underline{\bm{\theta}}}[ \log \mathbb{P}(  \underline{\vec{y}}  | \underline{\vec{X}}; \underline{\bm{\theta}}) |  \underline{\vec{y}} ] = \frac{1}{m} \sum_{l=1}^{L} \sum_{i=1}^{m} \log \left\{ \sum_{k=1}^{K} \tau_{k}^{l} \mathbb{P}(\vec{y}^{(i)} | \vec{X}^{(i)}; \bm{\theta}_{k}^{(l)})  \right\}. $$

As practitioners often seek for parsimonious models to represent the clusters, we present a rule-of-thumb to identify the point where there is diminishing return by further increasing the number of clusters, and hence to avoid potential over-fitting. Define the relative difference (RD) in DIC as

$$RD(k) = \frac{DIC_{k} - DIC_{k-1}}{DIC_{k-1}}, k=2,3,\cdots. $$

We define the optimal number of clusters given by a pre-specified cut-off value $\epsilon$ as $k_{opt}(\epsilon) = \min_{k} \left\{k | RD(k) \geqslant \epsilon  \right\}$, based on the reasoning that the optimal number of clusters should be the first $k$ resulting in limited relative improvement in terms of DIC. Simulation studies in section \ref{sec:Simulation} show empirical evidence supporting that $\epsilon = -0.005$ can be a reasonable rule-of-thumb for selecting the number of clusters.

We note that having an ensemble of networks makes it possible to assess the out-of-sample performance of mixture of ERGMs using the traditional statistical principle of cross-validation (CV), and there is work on using CV to estimate the number of clusters for observations with continuous values \citep{fu2019estimating}. In particular, to reduce the possibility of accidentally dropping all graphs in a single cluster completely by holding out too many graphs simultaneously, leave-one-out CV should be favored. The loss function for the cross-validation procedure can be negative log-likelihood evaluated on the held-out data as well as prediction error with respect to any structural properties of interest (obtained by simulating from estimated model using training data). Though the CV is not Bayesian and violates the likelihood principle, it is easy to implement and obviates the need to choose a threshold for when to stop adding clusters based on the predictive power of the model. 

\subsection{Posterior probability of cluster membership}
An appealing aspect of mixture modeling is that the posterior probability of individuals belonging to each cluster (alternately: graphs having been generated by a particular process) can be conveniently obtained as

\begin{equation}
\label{eq:post_prob}
    \mathbb{P}( Z_{i} = k | \vec{y}^{(i)} ) = \int \frac{ \tau_{k}  \mathbb{P}(\vec{y}^{(i)} | \vec{X}^{(i)} ;\bm{\theta}_{k}) }{ \sum_{k=1}^{K} \tau_{k}  \mathbb{P}(\vec{y}^{(i)} | \vec{X}^{(i)}; \bm{\theta}_{k}) } \pi( \underline{\bm{\theta}}, \bm{\tau} | \underline{\vec{y}}) d\underline{\bm{\theta}}d\bm{\tau},
\end{equation}

where $\pi( \underline{\bm{\theta}}, \bm{\tau} | \underline{\vec{y}})$ is the posterior distribution of $ \underline{\bm{\theta}}, \bm{\tau}$. The integral \eqref{eq:post_prob} is computationally intractable. Hence we use posterior samples $\underline{\bm{\theta}}^{1}, \cdots, \underline{\bm{\theta}}^{L}$ and $\bm{\tau}^{1}, \cdots, \bm{\tau}^{L}$  to obtain its Monte-Carlo approximation, 

\begin{equation}
    \hat{\mathbb{P}}( Z_{i} = k | \vec{y}^{(i)} ) = \frac{1}{L} \sum_{l=1}^{L} \frac{ \tau_{k}^{l} \mathbb{P}(\vec{y}^{(i)} | \vec{X}^{(i)}; \bm{\theta}_{k}^{l} ) }{ \sum_{k=1}^{K} \tau_{k}^{l} \mathbb{P}(\vec{y}^{(i)} |\vec{X}^{(i)}; \bm{\theta}_{k}^{l} ) }.
\end{equation}

The posterior mode, i.e., $ \hat{Z}_{i} = \argmax_{k} \hat{\mathbb{P}}( Z_{i} = k | \vec{y}^{(i)} )$ can be used as the output for cluster analysis, provided that the goal is to obtain a deterministic cluster assignment.

\section{Simulation studies}
\label{sec:Simulation}
We conduct extensive simulation studies to show that the proposed approach is capable of selecting the true number of clusters, recovering the true cluster memberships and true model parameters. 

\subsection{Experimental settings}
The ground truth is available for the synthetic data, as we simulate networks from mixtures of ERGM distributions defined on three most commonly used network sufficient statistics but with distinct parameters,

\begin{itemize}
    \item  $ g_{1}(\vec{y}) = \sum_{i<j} y_{ij}$, total number of edges.
    \item  $ g_{2}(\vec{y}) = e^{\phi} \sum_{k=1}^{n-2} \left\{ 1 - (1-e^{-\phi})^{k}   \right\} EP_{k}(\vec{y})$, geometrically weighted edgewise shared partners (GWESP). Here $EP_{k}(\vec{y})$ is the number of connected pairs that have exactly $k$ common neighbors, which measures local clustering in a network. The decay parameter $\phi$ controls the relative contribution of $EP_{k}(\vec{y})$ to the GWESP statistic, and it is fixed at $0.25$ in this case. 
    \item $ g_{3}(\vec{y}; \vec{X}) = \sum_{i < j} y_{ij}\mathbbm{1}_{ \{\vec{X}_{i} = \vec{X}_{j}\} } $, total number of edges with endpoints sharing same value on nodal covariate $\vec{X}$, often known as nodematch term.
\end{itemize}

We fix nodal covariate $\vec{X}$ to be a binary variable, and let one half of nodes take value $0$, while the other half take value $1$ on $\vec{X}$. To examine the performance of the proposed approach across a range of different conditions, we run a full-factorial experiment on the following three treatments

\begin{itemize}
    \item Network size: 40, 100, 250.
    \item Number of clusters: 2, 3.
    \item Cluster size: 10, 20, 50.
\end{itemize}

\begin{figure}
\label{fig:simulated_networks}
\centering
\includegraphics[width=13.5cm]{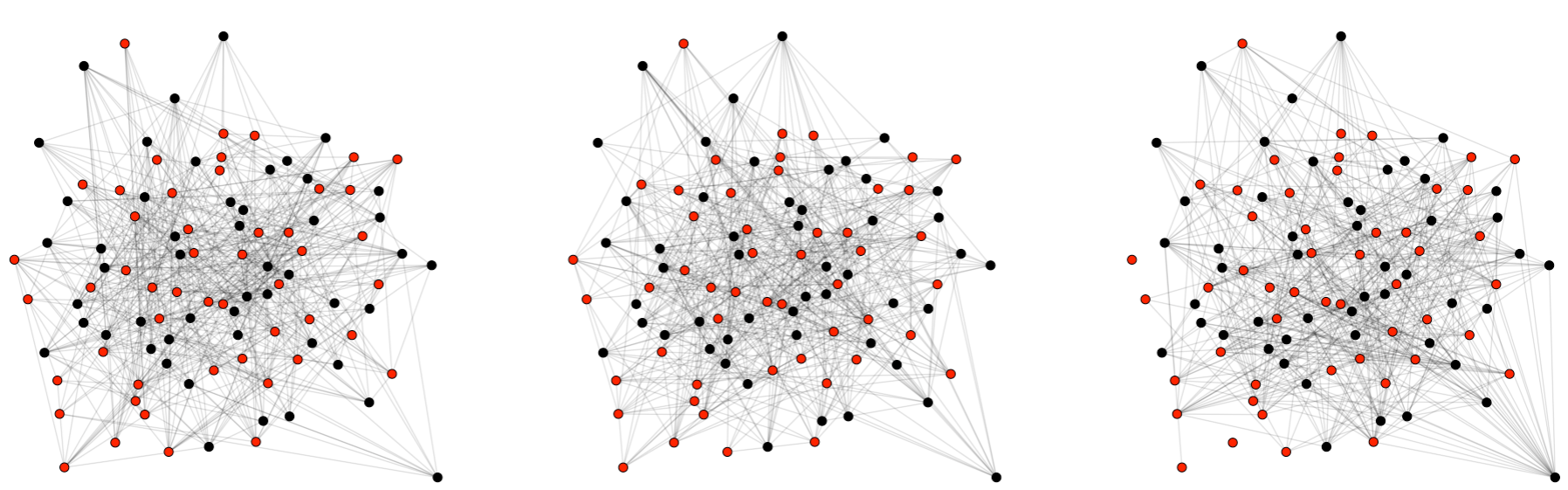}
\caption{Representative networks from clusters 1 (left), 2 (middle), and 3 (right). Network size: 100. Color indicates nodal covariate value: 0 (black), 1 (red). Despite the apparent similarity of the networks produced by the three generative processes, we are able to infer the latter from the observed ensemble. \label{fig:sim_networks}}
\end{figure}

 We thus have a total of 18 experimental conditions, each of which is run for 50 replicates. The true cluster-specific parameters are specified as


$$ \bm{\theta}^{40}_{true} = \begin{pmatrix}
   -1.15 & 0 & 0 \\
   -2.85 & 0.25 & 2.25 \\
   -4.95 & 2.5 & 0.25 \\ \end{pmatrix}, \ \ \bm{\theta}^{100}_{true} = \begin{pmatrix}
   -2.20 & 0 & 0 \\
   -4.15 & 0.25 & 2.25 \\
   -5.85 & 2.5 & 0.25 \\
   \end{pmatrix}, \ \ \bm{\theta}^{250}_{true} = \begin{pmatrix}
   -3.20 & 0 & 0 \\
   -4.95 & 0.25 & 2.25 \\
   -6.42 & 2.5 & 0.25 \\
   \end{pmatrix}$$
   
to ensure that the simulated networks (i) have similar mean degree ($\sim 9.9$, that is, networks of size 100 have density $\sim 0.10$) across different clusters and network sizes; and (ii) represent three most common-yet-intuitive patterns in real-world networks (parameter settings in the first row corresponds to the cases in which ties are independent Bernoulli draws, and parameter settings in the second row corresponds to the cases in which there is a strong homophily effect but a weak triadic closure effect, while the parameter settings in the third row correspond to the case in which there is a strong triadic closure effect but weak homophily effect). To maintain this pattern, we fix the values of coefficients associated with GWESP and nodematch terms across settings with different network sizes, and only modify the coefficient of edges term to keep the mean degree value as desired. We simulate networks using first two rows of the parameter matrices when the number of clusters is 2. Identifying subpopulations from ensembles of networks produced by this model is by no means a trivial task, especially as the cluster-specific parameters are chosen to produce networks of similar mean degrees ($\approx 0.10$) as shown in Figure \ref{fig:sim_networks}. While these networks appear superficially similar, we can recover the distinct processes that generated them.

We apply the proposed algorithm \ref{alg:alg1} to analyze the synthetic data sets, allowing the candidate values for the number of clusters to range from 1 to one greater than the true number of clusters (i.e., to 4, if the true number of clusters is 3; and  3, if the true number of clusters is 2). We assign random initial values to the latent indicator membership $\bm{Z}_{i}^{0}$, weight parameters $\bm{\tau}^{0}$ according to the prior, and set the parameters associated with the edge term as $-2$ (i.e., $\theta_{11} = \cdots = \theta_{K1} = -2$), while all other elements in $\underline{\bm{\theta}}$ are drawn independently from a uniform distribution $\mathcal{U}(-0.1,0.1)$. It is worth noting that our experiments suggest that better initial values can result in faster convergence and more stable performance for large networks. One effective way to initialize the proposed algorithm \ref{alg:alg1} is to first find the MPLE for each network in the ensemble separately, then cluster these MPLE estimates with K-means algorithm to initialize $\bm{Z}_{i}^{0}$ and calculate the intra-cluster mean MPLE estimates to determine the starting value of cluster-specific model parameters for each cluster. Table \ref{tb:sim_mcmc_setting} presents the MCMC settings, prior and proposal distribution for the experiments. The thinning interval is chosen as $50$ for all MCMC chains to obtain high-quality, weakly correlated draws from the posterior. All computations in this paper are implemented in \textbf{R} \citep{R2018}, and we use software suite \texttt{statnet} \citep{handcock2008statnet} to generate networks from ERGMs. 

\begin{table}[ht]
\centering
\caption{Total number of iterations, burn-in size, initialization method, prior hyper-parameters and covariance matrix for random-walk Metropolis-Hastings update of $\underline{\bm{\theta}}$ in simulation studies \label{tb:sim_mcmc_setting}}
\begin{tabular}{lllllll}
  \hline
 & Total iterations & Burn-in & Initialization & $\bm{\mu}$ & $\Psi$ & Prop. Cov \\ 
  \hline
40, 2 & 17500 & 7500 & Random  & (-1,0,0) & $25 I_{3}$ & $ 0.0025 I_{3}$ \\ 
40, 3 & 20000 & 10000 & Random & (-1,0,0)  & $25 I_{3}$ & $ 0.0025 I_{3}$ \\ 
100, 2 & 17500 & 7500 & Random & (-1,0,0) & $25 I_{3}$ & $ 0.0025 I_{3}$\\ 
100, 3 & 20000 & 10000 & MPLE, K-means  & (-1,0,0) & $25 I_{3}$ & $ 0.0025 I_{3}$\\ 
250, 2 & 22500 & 12500 & MPLE, K-means & (-1,0,0) & $25 I_{3}$  & $ 0.0016 I_{3}$ \\ 
250, 3 & 25000 & 15000 & MPLE, K-means & (-1,0,0) & $25 I_{3}$ & $ 0.0016 I_{3}$ \\ 
   \hline
\end{tabular}
\end{table}

\subsection{Recovery of true number of clusters and cluster membership}
\begin{figure}
\centering
\includegraphics[width=13.0cm, height=8cm]{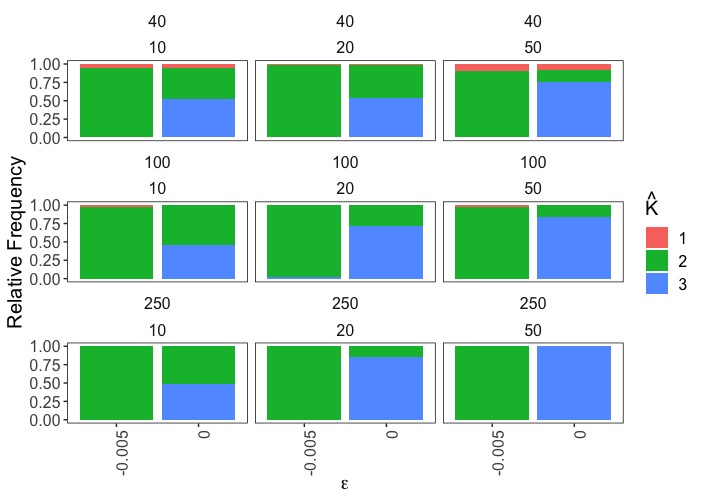}
\caption{Relative frequency of $\hat{K}$ selected by DIC criterion with $\epsilon = 0$ and $\epsilon = -0.005$. True number of clusters ($K$) = 2 \label{fig:RF_Khat_2}}
\end{figure}

\begin{figure}
\centering
\includegraphics[width=13.0cm, height=8cm]{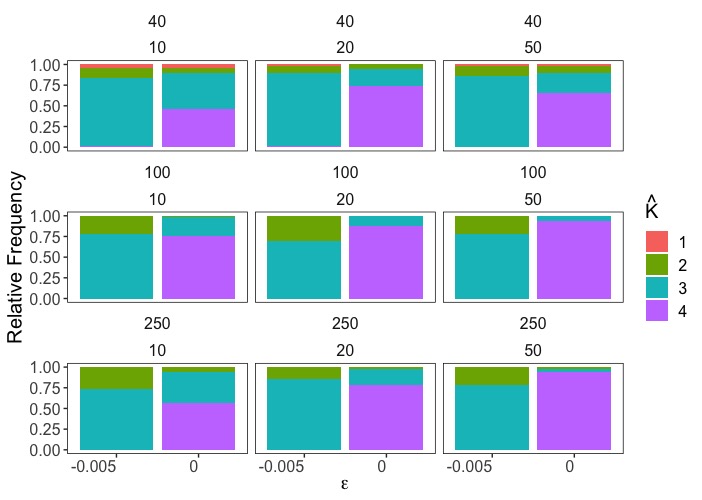}
\caption{Relative frequency of $\hat{K}$ selected by DIC criterion with $\epsilon = 0$ and $\epsilon = -0.005$. True number of clusters ($K$) = 3. \label{fig:RF_Khat_3}}
\end{figure}

We analyze the performance of proposed method in terms of its ability to identify the true number of clusters and cluster memberships. Figure \ref{fig:RF_Khat_2} and \ref{fig:RF_Khat_3} show that selecting the number of clusters according to the point beyond which there is diminishing return ($\epsilon = -0.005$) is unanimously superior to the minimum DIC criterion ($\epsilon = 0$), as the latter tends to be in favor of more complex models (i.e., with more clusters) than is optimal. Under DIC criterion with $\epsilon = -0.005$, we note that one has an $90\%$ or higher chance of identifying the true number of clusters when the true number is 2, and such chance is about $80\%$ when the true number of clusters is 3. Compared to identifying true number of clusters, recovering cluster memberships can be a more meaningful task in real-world applications, which we evaluate using adjusted rand index (ARI) \citep{hubert1985comparing}, a corrected-for-chance measure of the similarity between two clustering assignments, which yields a value of 1 for perfect cluster assignments and has an expected value of 0 for completely random cluster assignments. ARI is employed as an accuracy measure for cluster assignments here because the ground truth is available in the simulation study. Table \ref{tb:ARI} gives the mean ARI calculated across 50 replicates within each experiment setting, it shows that the proposed method can work well on the task of cluster assignments as all the mean ARI values are higher than 0.90 when the true number of clusters is $2$ and 0.85 when the true number of clusters is $3$ (a rule-of-thumb threshold value for ``good clustering'' is 0.80). We note that the mean ARI scores in Table \ref{tb:ARI} includes those calculated on the runs in which the true number of clusters is falsely identified, indicating that the proposed method is robust. In other words, the method fails gracefully, as it tends to completely combine two clusters or split one entire cluster into two when it errs, rather than mixing two clusters. 

\begin{table}[ht]
\centering
\caption{Mean ARI calculated across 50 replicates within each experiment setting. The true number of clusters is denoted as $K$. \label{tb:ARI}}
\begin{tabular}{l|lll|lll}
\hline
& & K=2 &  &  & K=3 & \\
  \hline
 & 10 & 20 & 50 & 10 & 20 & 50 \\ 
  \hline
40 & 0.940 & 0.980 & 0.900 & 0.902 & 0.942 & 0.924 \\ 
  100 & 0.980 & 0.996 & 0.980 & 0.902 & 0.869 & 0.905 \\ 
  250 & 1.000 & 1.000 & 1.000 & 0.884 & 0.939 & 0.905 \\ 
   \hline
\end{tabular}
\end{table}



\subsection{Estimation accuracy}
Given a correctly identified number of clusters, one natural question to ask is whether the proposed algorithm can accurately estimate the cluster-specific parameters. Specifically, we evaluate the estimation accuracy by examining the bias of posterior means. 

Table \ref{tb:bias_summary} summarizes the bias for cluster-specific model parameters under all experimental settings. We notice that the bias is in general small, especially for large networks, though there is slightly higher bias when the true number of clusters is 3. Large bias is mostly seen in the clusters in which there is strong dyadic dependence among edge variables (i.e., large coefficients associated with gwesp term), as expected. However, such bias becomes smaller and also less variable as sample size increases, indicating that larger sample size can mitigate the bias induced by the adoption of pseudo-likelihood. These findings offer implications to practitioners as estimated parameters are more reliable when large sample size is available or when the size of networks of interests is large.



\begingroup
\setlength{\tabcolsep}{11pt} 
\renewcommand{\arraystretch}{2.25} 
\begin{sidewaystable}
\centering
\caption{Mean (standard deviation) of bias across replicates in which the true number of clusters is correctly identified by DIC criterion ($\epsilon = -0.005$) within each experimental setting. \label{tb:bias_summary} }
\resizebox{\textwidth}{!}{
\begin{tabular}{l|llllll|lllllllll}
\toprule
&  &  & K=2  &  &  &  &  &  &  &  & K=3  &  &  &  & \\
  \hline
size & edges & gwesp & nodematch & edges & gwesp & nodematch & edges & gwesp & nodematch & edges & gwesp & nodematch & edges & gwesp & nodematch \\
40 & -2.85 & 0.25 & 2.25 & -1.15  & 0  & 0 & -4.95 & 2.5 & 0.25 & -2.85 & 0.25 & 2.25 & -1.15  & 0 & 0 \\
\midrule
\quad 10 & -0.041 (0.132) & 0.025 (0.087) & 0.01 (0.073) & 0.002 (0.095) & 0 (0.052) & -0.004 (0.047) & -0.06 (0.358) & 0.051 (0.25) & -0.014 (0.049) & 0.009 (0.153) & 0.007 (0.107) & -0.017 (0.07) & 0.02 (0.114) & -0.009 (0.058) & -0.005 (0.058) \\ 
\quad 20 & -0.014 (0.128) & 0.002 (0.078) & 0.011 (0.05) & -0.001 (0.069) & 0.002 (0.034) & -0.001 (0.034) & -0.021 (0.256) & 0.02 (0.174) & -0.014 (0.034) & 0.016 (0.121) & -0.004 (0.078) & -0.009 (0.051) & 0.002 (0.077) & 0.005 (0.042) & -0.008 (0.042) \\ 
\quad 50 & -0.003 (0.077) & -0.001 (0.049) & 0.006 (0.039) & 0.005 (0.046) & -0.003 (0.025) & 0.002 (0.02) & -0.046 (0.154) & 0.032 (0.109) & -0.006 (0.022) & 0.003 (0.069) & 0 (0.043) & -0.003 (0.033) & 0.006 (0.053) & -0.003 (0.029) & 0.001 (0.028) \\ 
\midrule
100 & -4.15 & 0.25 & 2.25  & -2.20  & 0 & 0  & -5.85  & 2.5 & 0.25 & -4.15 & 0.25 & 2.25 & -2.20 & 0 & 0 \\
\midrule
\quad 10 & -0.003 (0.048) & 0.003 (0.027) & -0.001 (0.051) & 0.01 (0.033) & -0.003 (0.018) & 0 (0.03) & 0.001 (0.107) & 0.002 (0.075) & 0.002 (0.035) & -0.007 (0.049) & 0 (0.029) & 0.006 (0.057) & 0.002 (0.044) & -0.001 (0.021) & 0.003 (0.026) \\ 
\quad 20 & 0.004 (0.034) & -0.002 (0.021) & -0.002 (0.036) & 0.008 (0.026) & -0.003 (0.013) & -0.001 (0.021) & -0.005 (0.081) & 0.006 (0.056) & -0.004 (0.029) & -0.005 (0.042) & 0.002 (0.023) & 0 (0.037) & -0.006 (0.03) & 0.003 (0.015) & 0.001 (0.022) \\ 
\quad 50 & 0.005 (0.02) & -0.004 (0.015) & 0.001 (0.027) & 0.002 (0.017) & 0 (0.007) & -0.001 (0.015) & -0.019 (0.055) & 0.013 (0.037) & 0 (0.017) & 0.003 (0.022) & -0.001 (0.014) & -0.002 (0.02) & -0.004 (0.015) & 0.001 (0.007) & 0 (0.016) \\ 
\midrule
250 & -4.95 & 0.25 & 2.25 & -3.20 & 0 & 0 & -6.42 & 0.25 & 2.25 & -4.95 & 0.25 & 2.25 & -6.42 & 2.5 & 0.25 \\
\midrule
\quad 10 & 0.004 (0.027) & -0.002 (0.011) & -0.001 (0.027) & -0.002 (0.015) & 0 (0.012) & 0 (0.019) & -0.012 (0.056) & 0.009 (0.042) & -0.001 (0.026) & 0.004 (0.025) & -0.001 (0.012) & -0.001 (0.028) & 0.001 (0.013) & 0.001 (0.012) & -0.006 (0.022) \\ 
\quad 20 & -0.003 (0.019) & 0 (0.008) & 0.003 (0.021) & 0.001 (0.01) & -0.001 (0.009) & 0 (0.011) & -0.009 (0.046) & 0.006 (0.032) & 0.001 (0.017) & -0.001 (0.018) & 0.001 (0.009) & -0.001 (0.021) & -0.001 (0.012) & 0.001 (0.007) & -0.002 (0.013) \\ 
\quad 50 & 0.001 (0.012) & 0 (0.006) & 0 (0.013) & 0 (0.007) & 0 (0.006) & -0.001 (0.007) & -0.003 (0.028) & 0.003 (0.019) & -0.002 (0.009) & 0 (0.014) & 0 (0.005) & 0.001 (0.015) & -0.001 (0.007) & 0 (0.005) & -0.002 (0.008) \\ 
   \hline
\end{tabular}}
\end{sidewaystable}
\endgroup

\subsection{Posterior predictive assessments}
One of the most appealing aspects of mixture modeling framework is that one can use simple probability distributions as building blocks to approximate complex probability distributions (e.g., mixtures of Gaussians are often used to approximate multimodal distributions). It is of substantial interest to see whether mixtures of ERGMs can provide an adequate fit to complex graph distributions. Although the selection of metrics should be guided by the particular properties of interests in practice, we consider four widely used metrics that characterize different aspects of graph structure as follows

\begin{itemize}
    \item Mean eigenvector centrality: the eigenvector centrality (EC) is a node-level metric that measures the degree of membership of a given node in the largest core/periphery structure in the graph, and we take mean eigenvector centrality among all nodes in the graph to convert it to a graph-level metric.\footnote{Except in very rare cases for which the graph adjacency matrix lacks a principal eigenvalue. In such circumstances, eigenvector centrality is a signed indicator of membership in the two largest core/periphery structures (positive versus negative).} The eigenvector centrality is also the best one-dimensional approximation of the graph structure (in a least-squares sense), and accuracy in reproducing it indicates the extent to which the model is able to recover the broadest structural features of the graph. 
    \item Transitivity: a standard measure of triadic closure in network analysis \citep{wasserman1994social}, defined as the ratio of complete triangles to all potentially complete triangles.
    \item Standard deviation of degree distribution: a measure of the level of heterogeneity in degree distribution.
    \item Mean of inverse geodesic distances: a measure of the overall closeness between nodes in a graph. 
\end{itemize}

\begin{figure}
\centering
\includegraphics[width=12.0cm, height = 6.5cm]{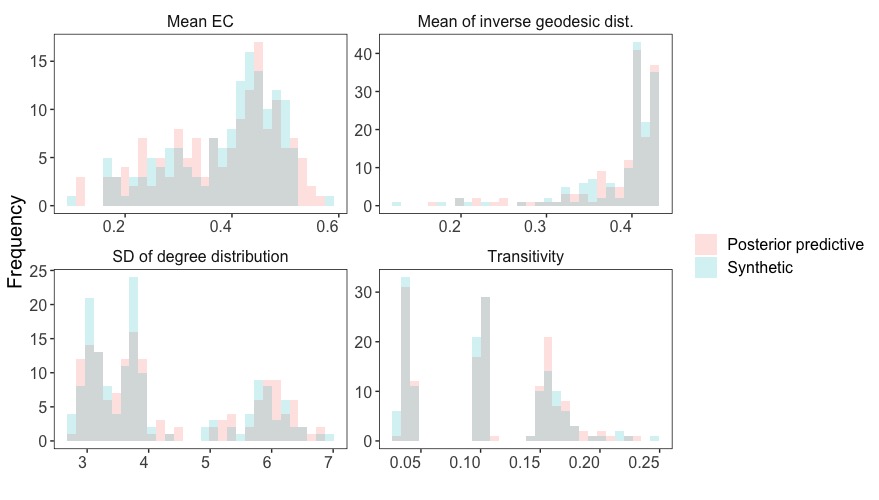}
\caption{Distribution of metrics of interests for posterior predictive samples and synthetic data, with corresponding Hellinger distance values : 0.150 (upper left), 0.283 (upper right), 0.141 (lower left), 0.076 (lower right).  \label{fig:post_check_1}}
\end{figure}

\begin{figure}
\centering
\includegraphics[width=12.0cm, height=6.5cm]{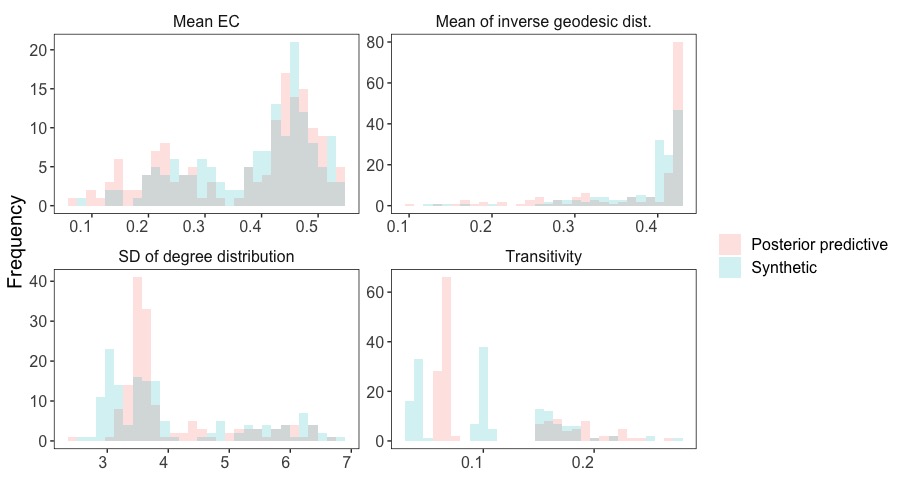}
\caption{Distribution of metrics of interests for posterior predictive samples and synthetic data, with corresponding Hellinger distance values: 0.173 (upper left), 0.270 (upper right), 0.125 (lower left), 0.105 (lower right).  \label{fig:post_check_2}}
\end{figure}

We focus on the experimental settings in which we have the most observations (3 clusters, 50 networks in each cluster) in this section. As each ensemble of networks in the synthetic data sets contains a total of 150 graphs, we also generate 150 networks using posterior samples with the data generating mechanism described in Figure \ref{fig:mixture_model_graph}. The simulated networks based on posterior samples and those synthetic networks are summarized by the four graph-level metrics, and their discrepancies are quantified in terms of the Hellinger distance, a commonly used metric for quantifying the distance between two probability distributions. We use function \texttt{CalcHellingerDist} in package \texttt{textmineR} \citep{textmineR} to calculate the empirical Hellinger distance between two sample vectors. 
Table \ref{tb:hd_summary} summarizes the mean and standard deviation of Hellinger distance evaluated across all replicates, regardless of whether the number of clusters selected by DIC criterion ($\epsilon = -0.005$) under the experimental settings of interests (i.e., true number of clusters is 3) is correct. The discrepancy between posterior predictive samples and synthetic data sets increases as the model selection accuracy decreases, from network size 40 to 100 and then to 250. To better understand the connections between Hellinger distance values and underlying visual difference in distributions in terms of histograms, we consider two representative replicates when the network size is 250. Figure \ref{fig:post_check_1} corresponds to a case in which the true number of clusters is selected and with Hellinger distance close to the average -- it is clear that the posterior predictive distribution of metrics of interest is very close to that of synthetic data. Figure \ref{fig:post_check_2} corresponds to a representative case in which the number of clusters is underestimated to be 2 -- the key observation is the resulting mixture model successfully captures the bimodal feature of mean eigenvector centrality and the left-skewed feature of mean of inverse geodesic distribution, and also identifies two of the three modes for standard deviation of degree distribution and transitivity. Although the result does not seem to be ideal, one key observation is that the resulting mixture model converges to the ``middle ground'' between two clusters, indicating that the possible reason for the model to choose two clusters over three is that the algorithm gets stuck at a local optimum, which might be mitigated by running MCMC chains longer or a more efficient proposal distribution for the Metropolis-Hastings step of Algorithm \ref{alg:alg1}. At a higher level, these results suggest the potential of mixtures of ERGMs as a tool to approximate complex graph distributions as one can view ERGMs as an analogue to ``kernel'' in density estimation.


\begin{table}[ht]
\centering
\caption{Mean (standard deviation) of Hellinger distance \label{tb:hd_summary}}
\begin{tabular}{lllll}
  \hline
 & Mean EC & Transitivity & SD of deg. dist. & Mean of inverse geodesic distance \\ 
  \hline
40, 3 & 0.045 (0.002) & 0.086 (0.007) & 0.083 (0.006) & 0.011 (0.001) \\ 
100, 3 & 0.076 (0.005) & 0.154 (0.009) & 0.123 (0.006) & 0.025 (0.003) \\ 
250, 3 & 0.145 (0.015) & 0.271 (0.016) & 0.137 (0.010) & 0.074 (0.016) \\ 
   \hline
\end{tabular}
\end{table}

\section{Case study}
\label{sec:Case_study}
In this section, we apply the proposed method to cluster the co-voting patterns among U.S. Senators from 1867 (start year of Congress 40) to 2014 (end year of Congress 113), which was a subset of the data first analyzed by \citet{moody2013portrait} using modularity and role-based blockmodels. The co-voting tendencies are represented by networks based on the roll call voting data from \url{http://voteview.com}, which contains the voting decision of each Senator (yay, nay, or abstain) for every bill brought to Congress \footnote{The data is available online in the R package \texttt{VCERRGM}, \url{https://github.com/jihuilee/VCERGM} }. The nodes in the co-voting network represent Senators and an edge is placed between two nodes if the corresponding Senators vote concurrently (both yay of both nay) on at least $75\%$ of the bills to which they were both present. Here we aim at identifying subgroups of networks that appear to have similar generating characteristics within the group but different characteristics across groups. 

\subsection{Model specification and estimation}
\begin{figure}
\centering
\includegraphics[width=13.5cm]{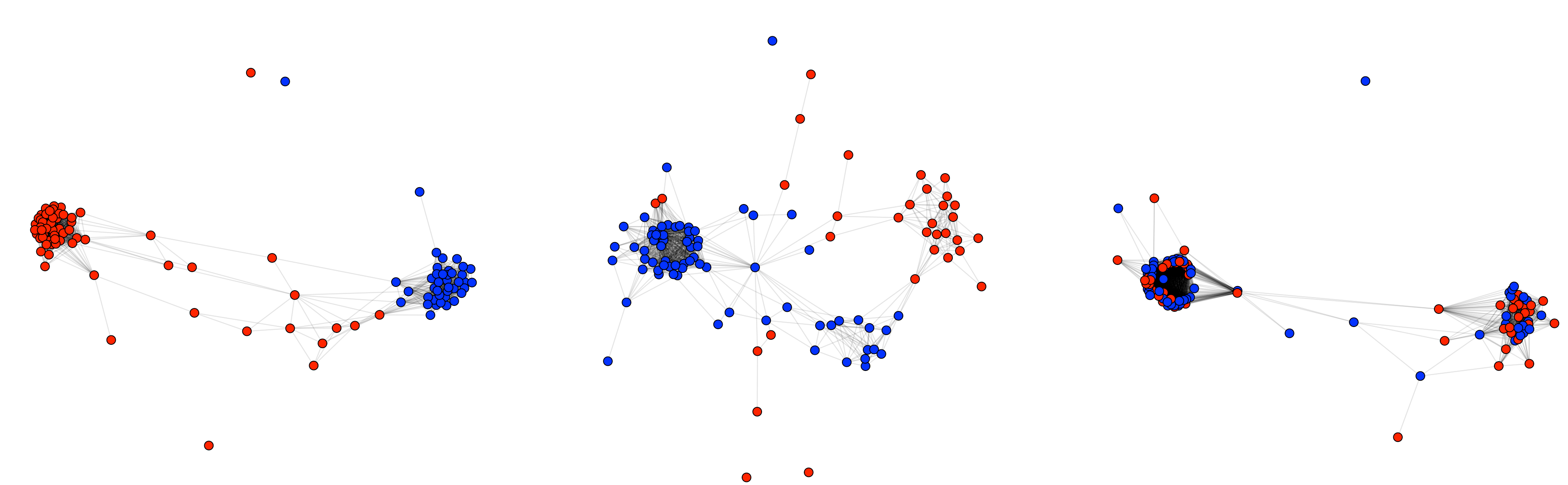}
\caption{Co-voting networks of 61st, 89th and 111th Congress, which were formed in the year of 1909, 1965 and 2009, respectively. Colors indicate Senators' party affiliations, blue = Democrats(D), red = Republican(R). \label{fig:congress}}
\end{figure}

Figure \ref{fig:congress} shows that the co-voting networks vary in structure on different years, and the party-affiliation appears to be a key factor affecting the co-voting patterns among Senators. Therefore we consider an ERGM model with following sufficient statistics 


$$ g_{1}(\vec{y}) = \sum_{i<j}y_{ij}, \ \text{total number of edges}; $$
$$ g_{2}(\vec{y}; \vec{X}) = \sum_{i<j}y_{ij}, \mathbbm{1}_{ \{\vec{X}_{i} = \vec{X}_{j} = D\} }, \ \ \text{total number of edges between Democrats};  $$
$$ g_{3}(\vec{y}; \vec{X}) = \sum_{i<j}y_{ij} \mathbbm{1}_{ \{\vec{X}_{i} = R, \vec{X}_{j} = D\} }, 
\ \ \text{total number of edges between Democrats and Republicans};$$
$$ g_{4}(\vec{y}) = e^{\phi} \sum_{k=1}^{n-2} \left\{ 1 - (1-e^{-\phi})^{k}   \right\} EP_{k}(\vec{y}), \ \text{GWESP statistic} $$

The decay parameter of GWESP term is fixed as $\phi=0.25$ as often used in ERGM literature. We note that these networks vary in size (range: $69-112$) and thus include an offset term \eqref{eq:krivitsky_offset} to adjust for network size. (This is equivalent to using the Krivitsky reference measure, which provides a parameterization with constant baseline expected degree.) We use the prior specification in \ref{sec:Mixture_of_ERGMs}, and run long MCMC chains (total iterations = 80000, burn-in = 30000, thinning interval = 50) with random initial values. 

\begin{figure}
\centering
\includegraphics[height=5cm, width=6.5cm]{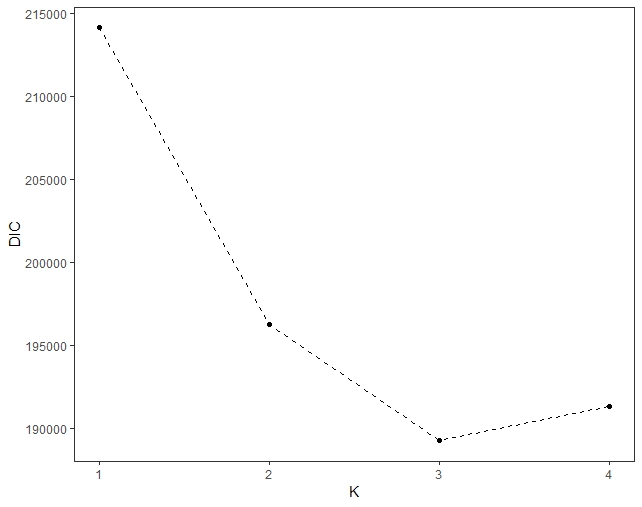}
\caption{DIC vs Number of clusters, Congress co-voting networks \label{fig:case_study_DIC}}
\end{figure}

\begin{figure}
\centering
\includegraphics[width=13.5cm]{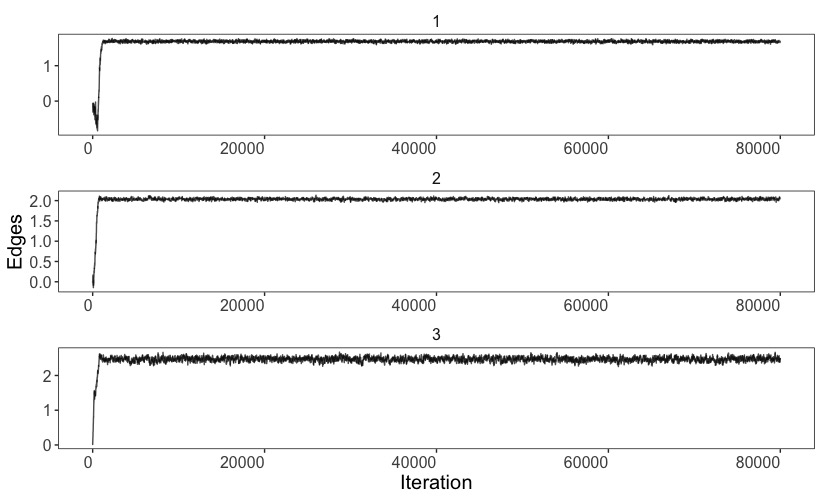}
\caption{Traceplots for parameters associated with edges term for 3 clusters. \label{fig:edges_trace}}
\end{figure}

Figure \ref{fig:case_study_DIC} indicates that the DIC reaches its minimum at $K=3$, and hence $K=3$ appears to be a plausible choice for the number of clusters. Under $K=3$, visual inspections on the traceplots suggest that the chains adapt to the high density region very fast and mix well (see Figure \ref{fig:edges_trace} for traceplots of edges parameter; other traceplots also show similar pattern, but are omitted in the interest of space). The posterior mean estimates of cluster-specific parameters are

$$ \hat{\bm{\tau}} = \begin{pmatrix} 
      0.36 \\
      0.47 \\
      0.17 \\
  \end{pmatrix}   \underline{\hat{\bm{\theta}}} = \begin{pmatrix}
  1.69 & 0.01 & -2.49 & 1.42 \\
  2.04 & -0.12 & -3.09 & 2.14 \\
  2.47 & 0.92 & -4.47 & 2.63 \\ \end{pmatrix} $$

We note that the size-invariant parameters for edge term (first column) can be interpreted as the log of the baseline mean degree (rather than the logit of the baseline density, as in the case of the counting measure), suggesting expected degrees varying from approximately 5.5 to 12 across clusters prior to consideration of other effects.

Based on these estimates, we have the following observations regarding the co-voting patterns.  Across all clusters, we see both inhibition of cross-party ties (third column) and strong triadic closure (fourth column).  Clusters do differ, however.  Cluster 1 shows essentially symmetric behavior by party (column two), with lower levels of cross-group inhibition and triadic closure bias than in the other clusters; overall, cluster 1 suggests a relatively low level of polarization by party, with voting only loosely restricted by party lines.  By contrast, cluster 2 reflects a much more polarized regime, with more activity overall and co-voting being more concentrated within party.  Like cluster 1, however, cluster 2 shows little party asymmetry (apart from a fairly weak tendency towards lower levels of co-voting among Democrats).  Such asymmetry is much more strongly pronounced within cluster 3, with intraparty Democratic ties being approximately 2.5 times as likely (ceteris paribus) as ties within the GOP. This cluster also reflects extremely high levels of polarization, with cross-party co-voting being strongly inhibited and high levels of triadic closure.  Over the period studied here, the most common pattern (probability 0.47) is the symmetric polarization of cluster 2, with the loose, low polarization pattern of cluster 1 also being fairly common (probability 0.36).  The asymmetric, highly polarized regime of cluster 3 is less common, but is still estimated to account for approximately $17\%$ of the observed cases.  Interestingly, we do not see a corresponding asymmetric pattern in which the GOP shows high intraparty vote density, as might be anticipated; thus, there appear to be latent differences in how the two parties behave during the period that, while not manifest in every congress, always have the potential to arise. 

One advantage of working with a fully generative model is the ability to perform ``what-if'' analyses that separate effects due to observed covariates from differences in structure arising from differences in generative processes.  To probe the impact of the three behavioral regimes inferred from the co-voting data, we consider how the entire ensemble Congressional networks would be expected to have been different, \emph{if} each respective regime had governed the U.S. Congress for the entire study period.  To perform such an analysis, we first simulate a set of posterior predictive networks for each Congress during the study period, with parameters drawn from the posterior distribution of each respective cluster.  Each collection of networks can be thought of as a simulated ``alternate history,'' in which the size and composition of each Congress were held to their real-world values but the behavioral tendencies that shaped the co-voting networks throughout the period were reflective of only one of the three clusters.  Systematic differences in network structure across sets thus provide insight into the potential impact of behavioral regime, controlling for size and composition.

\begin{figure}
\centering
\includegraphics[width=13.5cm]{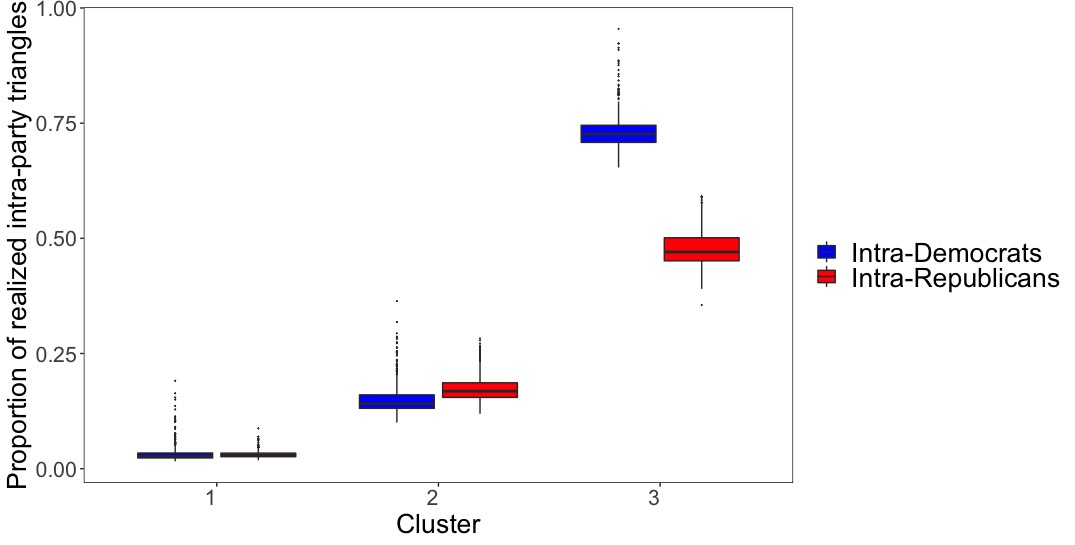}
\caption{Proportion of realized intra-party triangles in simulated networks. Colors indicate the party affiliation (blue = Democratic (D), red = Republican (R)).  \label{fig:triangles_intra_party}}
\end{figure}

\begin{figure}
\centering
\includegraphics[width=13.5cm]{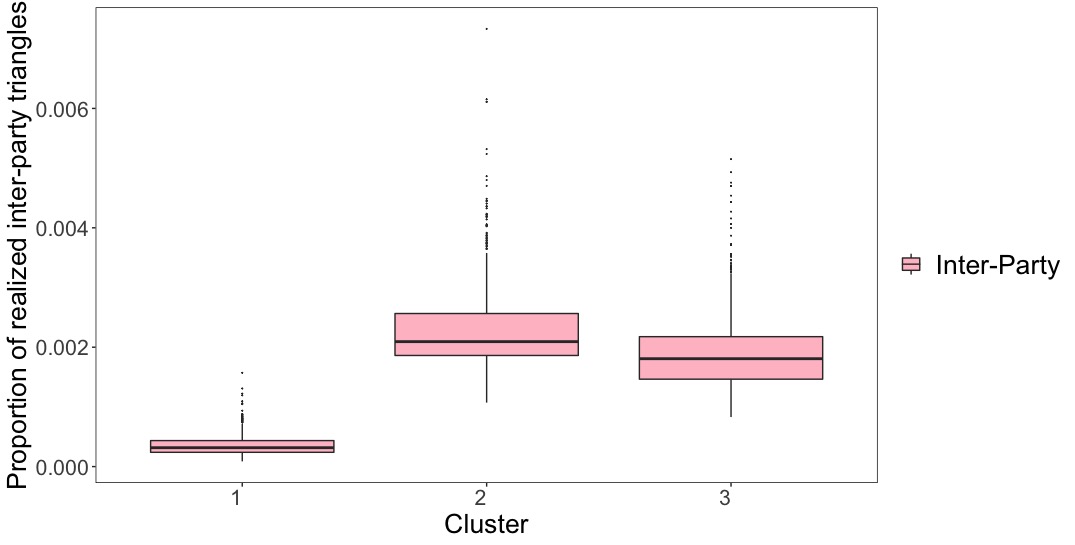}
\caption{Proportion of realized inter-party triangles in simulated networks. \label{fig:triangles_inter_party}}
\end{figure}

One important property that can be probed in this way is the expected incidence of voting coalitions, which play an important role in party politics.  Here, we focus on minimal coalitions, defined as sets of three legislators who consistently vote together (i.e., triangles).  Within-party coalitions can be sources of party cohesion, although they also act as blocks that can sometimes resist (and must be negotiated with by) party leaders; cross-party coalitions, by contrast, pose significant challenges to party cohesion, but can also serve as foci for sponsorship and promotion of bipartisan legislation.  Both are hence significant, with distinct implications for the political landscape.  To examine the coalition structures that would have been expected to occur under our three behavioral regimes, we simulate 10 ``alternate histories'' from the posterior distributions of each cluster, calculating the realized proportions of intra-Democratic, intra-Republican, and inter-Party triangles.  (That is, the counts of fully connected triads with all three members as Democrats, all three members as Republicans, or members from both parties, scaled by their maximum possible values.)  Using proportions rather than raw counts ensures these metrics are normalized for network size and the distribution of party affiliations in each Congress; substantively, this choice of scaling tells us how close each party (or the cross-party cut) is to forming a perfect coalition, in which all members vote in concert. Figure \ref{fig:triangles_intra_party} shows the realized proportion of intra-party triangles in simulated networks, and Figure \ref{fig:triangles_inter_party} shows the realized proportion of intra-party triangles in the simulated networks.  Both figures show substantial differences in coalition structure, implying that the behavioral regimes associated with the three inferred clusters would be expected to have a meaningful impact on the political process.  Specifically, we note the following:

\begin{itemize}
    \item The regime of cluster 1 is marked by the formation of very few voting coalitions, either within party or between party).  As suggested by the parameter values, we see little difference in coalition formation between the two parties, both having little cohesion.
    \item By contrast, the regime of cluster 2 shows a much higher incidence of intra-party coalition formation, with roughly 10-20\% of the potential intra-party coalitions being present.  Coalition incidence differs little by party, with at best a small average increment in the rate of coalition incidence for Republicans versus Democrats.  Interestingly, this regime also shows the highest rate of cross-party coalition formation; while the rate is very low overall, it is considerably higher than that observed under cluster 1.  
    \item Finally, the regime of cluster 3 favors extremely high levels of intra-party cohesion, with rates approaching 50\% of the maximum possible for Republicans and 75\% for Democrats.  As this implies, the resulting networks are also highly asymmetric, with the Democratic party expected to generate a much more cohesive coalition structure than the GOP.  Interestingly, this strong intra-party coalition formation does not exist entirely at the expense of cross-party coalitions: we find an expected rate of cross-party coalition formation that is only slightly less than that expected for networks arising under cluster 2.  That said, the much higher incidence of intra-party coalition formation under cluster 3 leads inter-party coalitions to be a smaller fraction of the total coalition set than under cluster 2, potentially making them less critical to the legislative process.
\end{itemize}

Taken together, these observations suggest that the cluster 1 regime tends to generate \emph{uniformly loose} voting networks with very few coalitions of any kind.  These networks may resist polarization, but their high level of fragmentation may make it more difficult to assemble the sorts of alliances needed to push through controversial legislation.  By contrast, the regime of cluster 2 tends to produce \emph{uniformly clustered} networks with moderately high levels of coalition formation in both parties coupled with relatively high numbers of cross-party coalitions.  These networks may pose particular challenges for party leaders, as they contain a mix of multiple local coalitions that must be courted for votes, ``lone wolves'' outside of coalitions who must be approached individually, and likely defectors whose cross-party coalitions provide a bullwark against within-party influence.  Finally, the regime of cluster 3 tends to produce \emph{party-cohesive} networks dominated by dense intra-party coalitions on both sides of the aisle (but with substantially higher levels of cohesion among Democratic legislators).  This regime offers party leaders the greatest chance of being able to mobilize members in support of legislation, at the cost of potential legislative deadlock during periods of high inter-party conflict.  

\begin{figure}
\centering
\includegraphics[width=13.5cm]{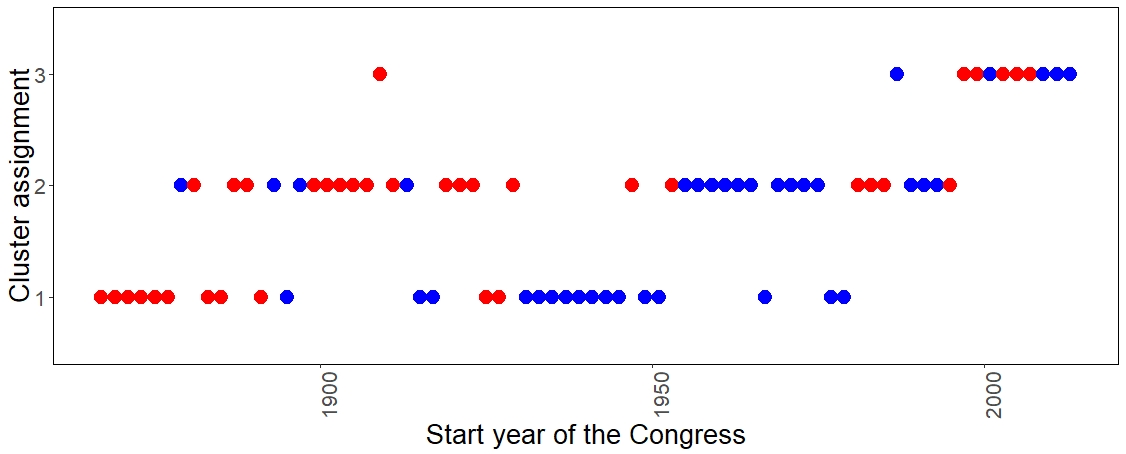}
\caption{Maximum probability cluster assignments over study period. Colors indicate the majority party in the corresponding Congress (blue = Democratic (D), red = Republican (R)). Regimes of voting behavior are visibly correlated over time.  \label{fig:cluster_label_year}}
\end{figure}

In addition to examining the potential impact of different behavioral regimes on voting networks, our model also provides insight into the incidence of these regimes over time.  For instance, Figure \ref{fig:cluster_label_year} shows maximum probability cluster assignments over the study period.  We see that the relatively symmetric cultures represented by cluster 1 and cluster 2 alternate in the nineteenth and twentieth centuries, while the culture of asymmetric polarization represented by cluster 3 becomes dominant after late 1990's. Such finding is in line with the current trend of political party polarization \citep{moody2013portrait}. Table \ref{tb:cluster_party} shows the  breakdown of congresses into $3 \times 2$ sub-categories according to the estimated co-voting pattern and the observed majority party. We examine the independence of co-voting pattern assignment and the majority party using Pearson's $\chi^2$ test, and we fail to reject the null hypothesis that the majority party is independent of the co-voting patterns ($\chi_{2}^{2} = 1.07$, p-value = $0.58$).  Thus, while the regimes of party behavior are quite visibly autocorrelated, this pattern does not seem to be related to which party has control of congress at any given time.

\begin{table}[ht]
\centering
\caption{Tabulation of co-voting pattern by majority party (from Congress 40 to Congress 113).  Majority party is not significantly related to voting regime. \label{tb:cluster_party}}
\begin{tabular}{l|ll}
  \hline
Co-voting Pattern & Democratic & Republican \\ 
  \hline
1 &  16 &  11 \\ 
  2 &  17 &  19 \\ 
  3 &   5 &   6 \\ 
   \hline
\end{tabular}
\end{table}

\subsection{Model assessment}
To assess the adequacy of the resulting model, we consider a simulation-based method \citet{hunter2008goodness}, with the basic insight that a fitted ERGM model should be able to reproduce in simulation structural properties similar to those of the observed networks. Instead of simulating from a single point estimate, we propose to simulate networks from estimated posterior distribution, following practices of posterior predictive assessment in the Bayesian literature \citep{gelman1996posterior}. The structural property of interest here is the modularity score \citep{newman2006modularity} (assessed by party), which can be interpreted as a measure of the polarization of networks with respect to party structure. By definition, the modularity score ranges from $-1$ to $1$, with larger values indicating higher levels of polarization. We replicate the following evaluation procedure for $100$ times:

\begin{enumerate}
    \item For each vertex set, we first randomly draw a latent membership indicator using posterior samples of $\bm{\tau}$, then simulate a network from the corresponding component using posterior samples of $\bm{\theta}$. 
    \item Compute the modularity score of the observed ensemble of networks and the simulated networks.
\end{enumerate}

We compare the distribution of modularity scores of simulated networks to that of observed networks using Hellinger distance. We obtain the mean of Hellinger distance values as $0.095$ and standard deviation of Hellinger distance values as $0.002$.

\begin{figure}
\centering
\includegraphics[width=9cm, height=6cm]{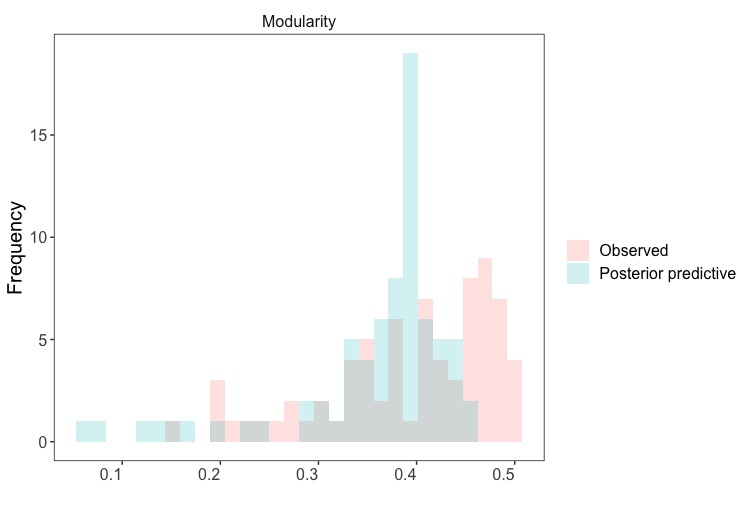}
\caption{Modularity scores of simulated and observed ensemble of networks. Hellinger distance: 0.096.\label{fig:case_study_gof}}
\end{figure}

Figure \ref{fig:case_study_gof} shows the distribution of modularity scores for a replicate that has average-case performance (Hellinger distance value : 0.096). We see that the resulting mixture model can capture not only the left-skewed feature of the modularity scores in the observed data but also the variation the observed modularity scores to a large extent. The remaining discrepancy between observed modularity scores and those of simulated networks might be mitigated by more accurate estimation algorithms for cluster-specific parameters (e.g., using importance sampling to approximate ERGM likelihood rather than the pseudo-likelihood), at higher computational expense.

\section{Conclusion}
\label{sec:Conclusion}
In this paper, we proposed a mixture of ERGMs approach for modeling the generative process leading to heterogeneous network ensembles. We developed a Metropolis-within-Gibbs algorithm to fit ERGM mixtures and obtained Bayesian estimates of clustering assignment probabilities and the cluster-specific ERGM parameters. To account for the difference in the size of the observed networks, we used a size-adjusted parameterization for ERGMs. We also tailored a version of observed DIC and defined an empirical rule to select the number of clusters, which is proved to be effective in simulation study. The simulation studies also showed that the proposed approach can accurately recover the cluster membership and cluster-specific parameters, without requiring much effort on initialization. 

We applied the proposed approach to study the political co-voting networks among U.S. Senators, and identified three clusters that represent vastly different co-voting patterns. After matching the clusters with temporal information, we observed that one symmetric co-voting pattern and another mildly asymmetric co-voting patterns alternate in nineteenth and twentieth century, and there appeared to be an abrupt shift in the co-voting pattern towards the direction of political party polarization in last two decades. 

Compared to other methods in the literature, our proposed method allows straightforward statistical inference for the generative processes of heterogeneous ensembles of networks with edgewise dependence, and is conveniently interpretable. We believe that the proposed method can prove to be a highly effective tool for both exploratory and inferential analysis of ensembles of networks. 

In closing, we comment on three important directions of future research that could prove beneficial to the modeling of ensembles of networks: the development of more sophisticated size-adjusted parameterizations, more accurate tractable approximations of the ERGM likelihood and Dirichlet Process mixtures of ERGMs. It is worth mentioning that the sizes of the US congresses between 1867 and 2014 range from 69 to 112, non-identical but broadly similar. More importantly, these size changes occur within a social system whose basic structure remains fairly similar throughout the time period.  In other cases, however, large size differences may be accompanied by increasingly complex internal barriers to interaction or other additional exogenous structure that must be accounted for to obtain realistic predictions.  While this additional structure is not available in the form of covariates, more sophisticated size-adjusted parameterizations may be required; reference measures or other tools facilitating ``automatic'' correction of such effects would facilitate mixture modeling in such scenarios. With respect to likelihood calculation, it is encouraging that we obtain favorable results in our simulation study using the easily computed pseudo-likelihood approximation.  In particular, the main deficiency of the pseudo-likelihood is excessive sharpness near the mode, which could in principle encourage the over-production of mixture components.  While we do not see this effect here, more accurate likelihood approximations that are inexpensive enough to perform at each MCMC step for large models would be desirable. As such improved approximations become available, they can be easily integrated into the posterior simulation framework described here. Last  but not least, a natural further extension of the finite mixture modeling framework could be Dirichlet Process mixtures of ERGMs where the number of mixture components can vary depending on the incoming data size. Although computationally challenging, such an extension can provide a highly flexible-yet-interpretable density estimation framework for complex graph distributions. 

\bigskip







\bibliographystyle{abbrvnat}
\bibliography{mixergm.bib}

\end{document}